\documentclass{SolarPhysics}    % Specifies the document style.
\usepackage[optionalrh]{spr-sola-addons}
\usepackage{epsfig}
\usepackage{url}

\newcommand{\gapprox}{\lower.4ex\hbox{$\;\buildrel >\over{\scriptstyle\sim}\;$}}
\newcommand{\lapprox}{\lower.4ex\hbox{$\;\buildrel <\over{\scriptstyle\sim}\;$}}

\def\etal{{\it et al.}~}
\def\eg{{\it e.g.}~}
\def\ie{{\it i.e.}~}
\def\etc{{\it etc.}~}
\begin{document}
\begin{article}
\begin{opening}
\title{ The State of Self-Organized Criticality of the Sun 
	During the Last Three Solar Cycles. I. Observations }

\author{Markus J. Aschwanden}
\runningauthor{M. Aschwanden}
\runningtitle{Self-Organized Criticality}

\institute{Solar and Astrophysics Laboratory,
	Lockheed Martin Advanced Technology Center, 
        Dept. ADBS, Bldg.252, 3251 Hanover St., Palo Alto, CA 94304, USA; 
        (e-mail: \url{aschwanden@lmsal.com})}

\date{Received 1 July 2010; Accepted ...}

\begin{abstract}
We analyze the occurrence frequency distributions of peak fluxes $P$, 
total fluxes $E$, and durations $T$ of solar flares over the last 
three solar cycles (during 1980--2010) from hard X-ray data of
HXRBS/SMM, BATSE/CGRO, and RHESSI. From the synthesized data 
we find powerlaw slopes with mean values of
$\alpha_P=1.72\pm0.08$ for the peak flux,
$\alpha_E=1.60\pm0.14$ for the total flux, and
$\alpha_T=1.98\pm0.35$ for flare durations. 
We find a systematic anti-correlation of the powerlaw slope of peak fluxes 
as a function of the solar cycle, varying with an
approximate sinusoidal variation $\alpha_P(t)=\alpha_0+\Delta \alpha
\cos{[2\pi (t-t_0)/T_{cycle}]}$, with a mean of $\alpha_0=1.73$, a variation
of $\Delta \alpha =0.14$, a solar cycle period $T_{cycle}=12.6$ yrs,
and a cycle minimum time $t_0=1984.1$.
The powerlaw slope is flattest during the maximum of a solar cycle,
which indicates a higher magnetic complexity of the solar corona
that leads to an overproportional rate of powerful flares.
\end{abstract}


\end{opening}

\section{       Introduction 	}

In the observational part of this study (Paper I) we focus on the statistics 
of solar-flare hard X-ray fluxes during the course of the last three solar 
cycles, while theoretical modeling is referred to Paper II. 
Solar flares are catastrophic events in the solar corona,
most likely caused by a magnetic instability that triggers a magnetic
reconnection process, producing emission in almost all wavelengths. 
Since the emission mechanisms are all different in each wavelength, such as
nonthermal bremsstrahlung (in hard X-rays), thermal bremsstrahlung 
and free-bound or recombination radiation (in
soft X-rays and EUV), gyrosynchrotron emission (in microwaves), plasma 
emission (in metric and decimetric waves), \etc , estimates of the energy 
contained in each flare event strongly depends on the emission mechanism, 
and thus on the wavelength. In this study we concentrate on the hard X-ray 
wavelength. Hard X-ray emission in solar flares mostly results from 
thick-target bremsstrahlung of nonthermal particles accelerated in the 
corona that precipitate into the dense chromosphere. Thus, the hard X-ray 
flux is the most direct measure of the energy release rate, and thus is 
expected to characterize the energy of flare events in a most uncontaminated 
way, while emission in other wavelengths exhibit a more convolved evolution
of secondary emission processes. The main observables that are available
for flare statistics in hard X-rays are the peak flux $P$, the total 
flux or fluence $E$ (defined as the time-integrated flux over the entire event),
and the total time duration $T$ of the event. In this study, we present
a comprehensive compilation of occurrence frequency distributions of these
observables obtained in hard X-rays, and investigate whether their behavior  
is different during solar cycle minima, including the current
anomalous solar minimum. 

\section{Statistics of Solar Flares in Hard X-rays}

In this Section we describe observed occurrence frequency distributions 
of solar flare hard X-ray parameters (in chronological order), 
discuss the properties of the datasets (hard X-ray energies, observational
epochs, instruments), compile the results in Table 1 (powerlaw slopes of
fluxes, fluences, durations, number of events, instruments, and 
references), and derive synthesized distributions that serve
as reference values of the average solar flare activity.
In a subsequent Section we investigate whether deviations from these
reference values can be found during the solar cycle. 

\begin{table}
\begin{center}
\footnotesize	
\caption{Frequency distributions measured from solar flares in hard X-rays
and gamma-rays. References: $^1$) Datlowe \etal(1974);
$^2)$ Lin \etal(1984); $^3)$ Dennis (1985); $^4)$ Schwartz \etal(1992);
$^5)$ Crosby \etal(1993); $^6)$ Biesecker \etal(1993);
$^7)$ Biesecker \etal(1994);
$^8)$ Crosby (1996); $^9)$ Lu \etal(1993); $^{10})$ Lee \etal(1993);
$^{11})$ Bromund \etal(1995); $^{12})$ Perez Enriquez and Miroshnichenko 
(1999); $^{13}$ Su \etal(2006); $^{14}$ Christe \etal(2008);
$^{15})$ Lin \etal(2001); $^{16})$ Tranquille \etal(2009).}
\begin{tabular}{lllrlr}
\hline
Powerlaw        &Powerlaw 	&Powerlaw   &Number     &Instrument&Reference\\
slope of	&slope of	&slope of   &of         &and	 &\\
peak flux	&fluence  	&durations  &events	&threshold&\\
$\alpha_P$ 	&$\alpha_E$	&$\alpha_T$ &$n$        &energy	 &\\
\hline
1.8		&		&	    &123        &OSO--7($>$20 keV)&1)\\
2.0		&		&	    &25         &UCB($>$20 keV)   &2)\\
1.8		&		&	    &6775       &HXRBS($>$20 keV) &3)\\
1.73$\pm$0.01   &               &           &12,500     &HXRBS($>$25 keV) &4)\\
1.73$\pm$0.01   &1.53$\pm$0.02  &2.17$\pm$0.05 &7045    &HXRBS($>$25 keV) &5)\\
1.71$\pm$0.04   &1.51$\pm$0.04  &1.95$\pm$0.09 &1008    &HXRBS($>$25 keV) &5)\\
1.68$\pm$0.07   &1.48$\pm$0.02  &2.22$\pm$0.13 &545     &HXRBS($>$25 keV) &5)\\
1.67$\pm$0.03   &1.53$\pm$0.02  &1.99$\pm$0.06 &3874    &HXRBS($>$25 keV) &5)\\
1.61$\pm$0.03   &               &	    &1263       &BATSE($>$25 keV) &4)\\
1.75$\pm$0.02   &               &	    &2156       &BATSE($>$25 keV) &6)\\	
1.79$\pm$0.04   &               &	    &1358       &BATSE($>$25 keV) &7)\\	
1.59$\pm$0.02   &               &2.28$\pm$0.08 &1546    &WATCH($>$10 keV) &8)\\
1.86		&1.51		&1.88	    &4356	&ISEE--3($>$25 keV) &9)\\
1.75		&1.62           &2.73       &4356       &ISEE--3($>$25 keV) &10)\\
1.86$\pm$0.01   &1.74$\pm$0.04  &2.40$\pm$0.04 &3468    &ISEE--3($>$25 keV) &11)\\
1.80$\pm$0.01	&1.39$\pm$0.01	&	    &110        &PHEBUS($>$100 keV) &12)\\
1.80$\pm$0.02	&               &2.2$\pm$1.4&2759       &RHESSI($>$12 keV)  &13)\\
1.58$\pm$0.02	&1.7$\pm$0.1    &2.2$\pm$0.2&4241       &RHESSI($>$12 keV)  &14)\\
1.6	        &               &           &243        &BATSE($>$8 keV)    &15)\\
1.61$\pm$0.04   &               &           &59         &ULYSSES($>$25 keV) &16)\\
\hline
\end{tabular}
\end{center}
\end{table}

\medskip
One of the earliest reports of a frequency distribution of solar hard
X-ray flare fluxes was made by Datlowe \etal(1974), who published
a frequency distribution of 123 flare events detected
in the 20--30 keV energy range above a threshold of $\gapprox 0.1$
photons (cm$^{-2}$ s$^{-1}$ keV$^{-1}$) with the OSO--7 spacecraft 
during the period of 10 October 1971 -- 6 June 1972, 
finding a powerlaw slope of $\beta_P
\approx 0.8$ for the {\sl cumulative frequency distribution}. 
For compatibility we list only powerlaw slopes $\alpha$ of 
{\sl differential frequency distributions} in Table 1, 
and use the conversion $\alpha = \beta + 1$ when needed.
We list also the number of events, which is a good indicator of the 
statistical uncertainty of the powerlaw slope fits (Figure 6).

A sample of 25 microflares of smaller size were detected at 20 keV with
a balloon-borne instrumentation of {\sl University of California Berkeley 
(UCB)} during 141 minutes of observations on
27 June 1980, yielding a powerlaw distribution with a slope of $\beta
\approx 1$ (Lin \etal1984). 

Many more events were observed with the
{\sl Hard X-Ray Burst Spectrometer (HXRBS)} onboard the {\sl Solar Maximum 
Mission (SMM)} spacecraft, which recorded 6775 flare events during the 
1980--1985 period, exhibiting a powerlaw distribution of peak count 
rates with a slope of $\alpha_P=1.8$, extending over four orders of magnitude 
(Dennis 1985).

A next mission with hard X-ray detector capabilities was
the {\sl Compton Gam\-ma Ray Observatory} (CGRO). Although it was designed
to detect gamma-rays from astrophysical objects, it also detected 
solar flares systematically during the period of 1991--2000.  
Using the {\sl Burst And Source Transient Experiment} (BATSE), 
statistics of flares with energies 
$>25$ keV was sampled and more detailed powerlaw distributions of peak
fluxes were reported with values of $\alpha_P=1.61\pm0.03$ 
(Schwartz \etal1992), $\alpha_P=1.75\pm0.02$ (Biesecker \etal1993),
and $\alpha_P=1.79\pm0.04$ (Biesecker \etal1994) for BATSE.   
Biesecker (1994) noticed slight differences of the powerlaw slope
during low activity $(\alpha_P=1.71\pm0.04$) and high activity periods
($\alpha_P=1.68\pm0.02$), with the powerlaw slope usually flatter for
high-activity periods. 

A systematic study of flares observed with HXRBS over the entire mission
duration of 1980--1989 was conducted by Crosby \etal(1993), measuring
peak count rates $P_{cts}$ (cts s$^{-1}$), converted into photon 
fluxes $P_{ph}$ (photons cm$^{-2}$ s$^{-1}$ keV$^{-1}$) at energies $>25$ keV, 
peak HXR spectrum-integrated fluxes $P_{X}$ (photons cm$^{-1}$ s$^{-1})$,
peak electron fluxes $P_{e}$ (ergs s$^{-1}$), flare durations $T$,
and time-integrated total energies in electrons $E_e$ (ergs), for four 
different time intervals of the solar cycle. In Table 1 we list the values 
for the time ranges of 1980--1982 (7045 events; solar maximum phase),
1983--1984 (1008 events), 1985-1987 (545 events; solar minimum phase),
and 1988-1989 (3874 events). The values of the powerlaw slopes change 
by $\lapprox 2\%$ during different periods of the solar cycle.

From the {\sl Wide Angle Telescope for Cosmic Hard X-Rays} (WATCH) onboard
the Russian satellite GRANAT, a sample of 1546 flare events was observed
at energies of 10--30 keV during 1990--1992, yielding similar powerlaw slopes
for peak count rates, $\alpha_P=1.59 \pm 0.02$, and flare durations $\alpha_T
=2.28\pm0.08$ as reported before (Crosby, 1996; Crosby \etal1998). 
However, it was noted that the frequency distribution of flare durations 
exhibits a gradual rollover for short flare durations, approaching a slope
of $\alpha_T \approx 1$, so it cannot be fitted with a single powerlaw
distribution over the entire range of flare durations. From the PHEBUS
instrument on GRANAT, which is sensitive to gamma-ray energies,
Perez Enriquez and Miroshnichenko (1999) analyzed 110 high-energy solar flares
observed in the energy range of 100 keV -- 100 MeV and found the following
powerlaw slopes:
$\alpha_P=1.80\pm0.01$ for (bremsstrahlung) hard X-ray fluxes at $>100$ keV,
$\alpha_P=1.38\pm0.01$ for photon energies at 0.075 -- 124 MeV,
$\alpha_P=1.39\pm0.01$ for bremsstrahlung at 300 -- 850 keV,
$\alpha_E=1.50\pm0.03$ for the 511 keV annihilation line fluence,
$\alpha_E=1.39\pm0.02$ for the 2.223 MeV neutron line fluence, and
$\alpha_E=1.31\pm0.01$ for the 1 -- 10 MeV gamma-ray line fluence. 
The flatter powerlaw slopes of the occurrence frequency distributions
at higher energies could possibly be explained by flatter spectra
(Section 3.1).

\begin{figure}
\centerline{\includegraphics[width=1.0\textwidth]{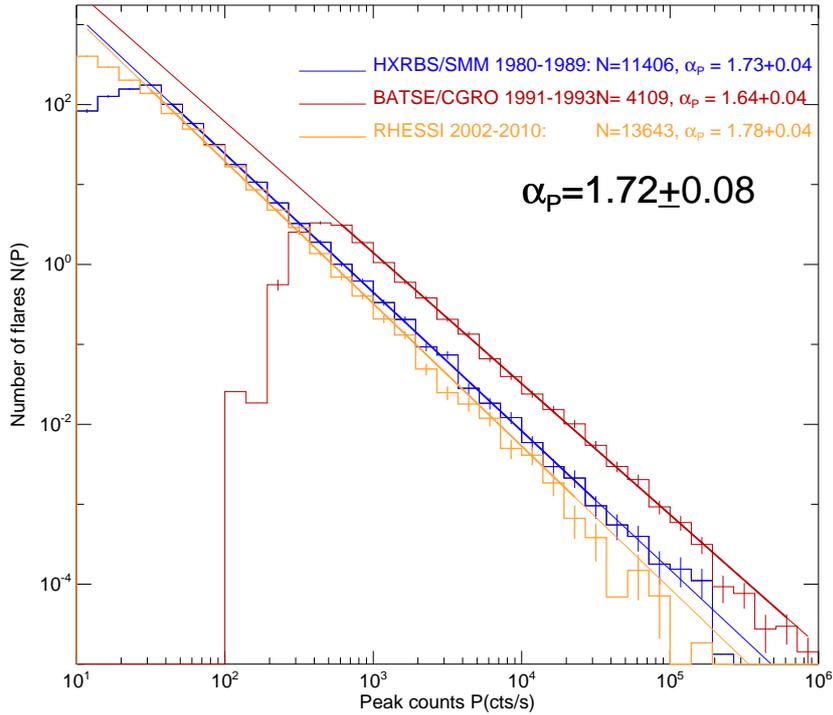}}
\caption{Occurrence frequency distributions of hard X-ray peak count rates
$P$ (cts/s) observed with HXRBS/SMM (1980 -- 1989), BATSE (1991 -- 1993),
and RHESSI (2002 -- 2010), with powerlaw fits. An average pre-flare 
background of $40$ cts s$^{-1}$ was subtracted from the HXRBS 
count rates. Note that BATSE/CGRO has larger detector areas, and thus 
records higher count rates.}
\end{figure}

\begin{figure}
\centerline{\includegraphics[width=1.0\textwidth]{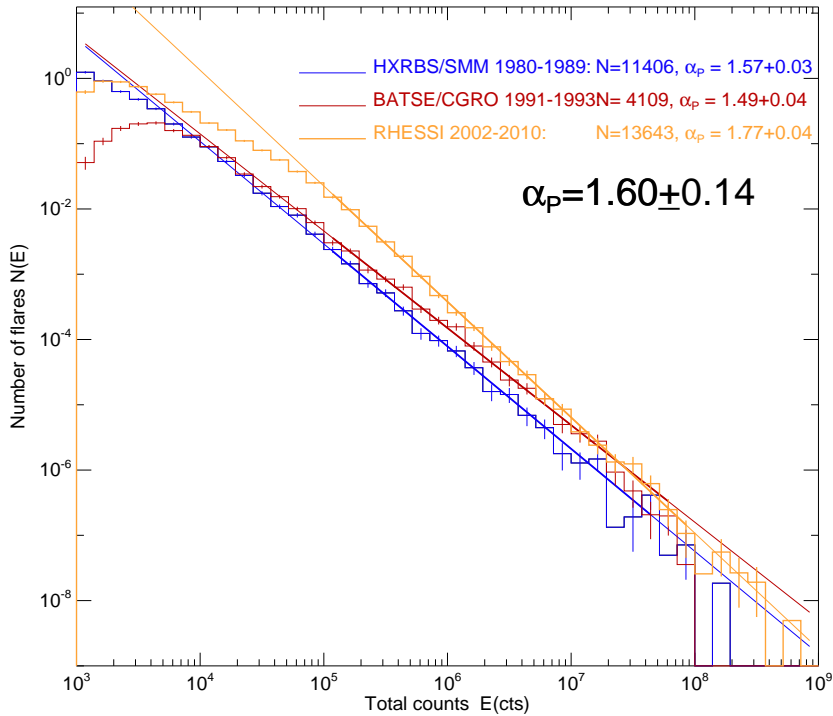}}
\caption{Occurrence frequency distributions of hard X-ray total counts
or fluence $E$ (cts) observed with HXRBS/SMM (1980 -- 1989), 
BATSE (1991 -- 1993), and RHESSI (2002 -- 2010), with powerlaw fits.
An average pre-flare background of $40$ cts s$^{-1}$ mulitplied with
the flare duration was subtracted in the total counts of HXRBS.}
\end{figure}

\begin{figure}
\centerline{\includegraphics[width=1.0\textwidth]{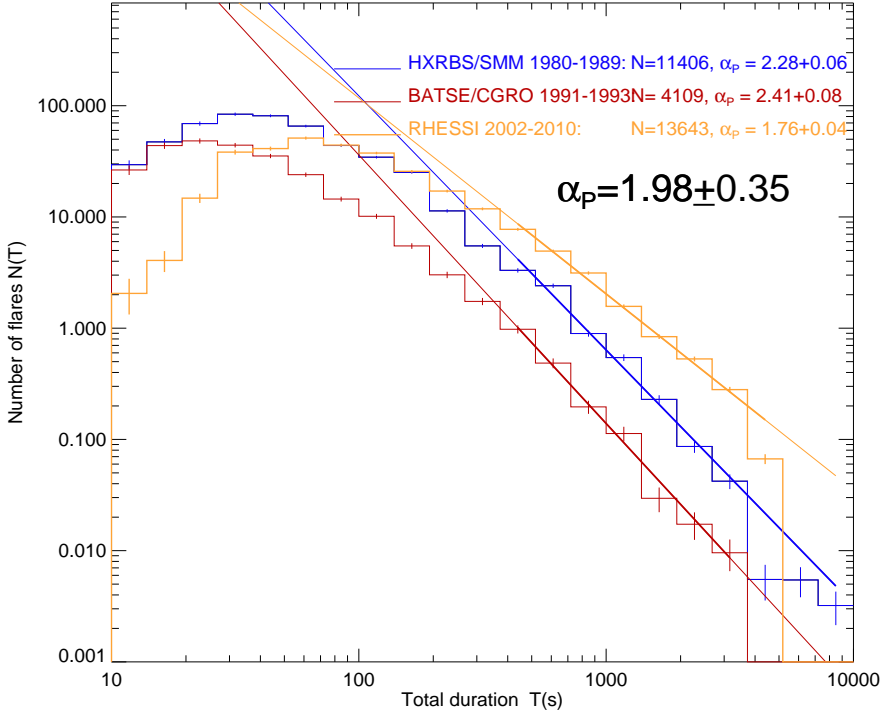}}
\caption{Occurrence frequency distributions of hard X-ray flare durations 
$T(s)$ observed with HXRBS/SMM (1980 -- 1989), BATSE (1991 -- 1993),
and RHESSI (2002 -- 2010), with powerlaw fits. The flare durations for RHESSI
were estimated from the time difference between the start and peak time,
because RHESSI flare durations were determined at a lower energy of 12 keV 
(compared with 25 keV for HXRBS and BATSE), where thermal emission prolonges
the nonthermal flare duration.}
\end{figure}

\begin{figure}
\centerline{\includegraphics[width=1.0\textwidth]{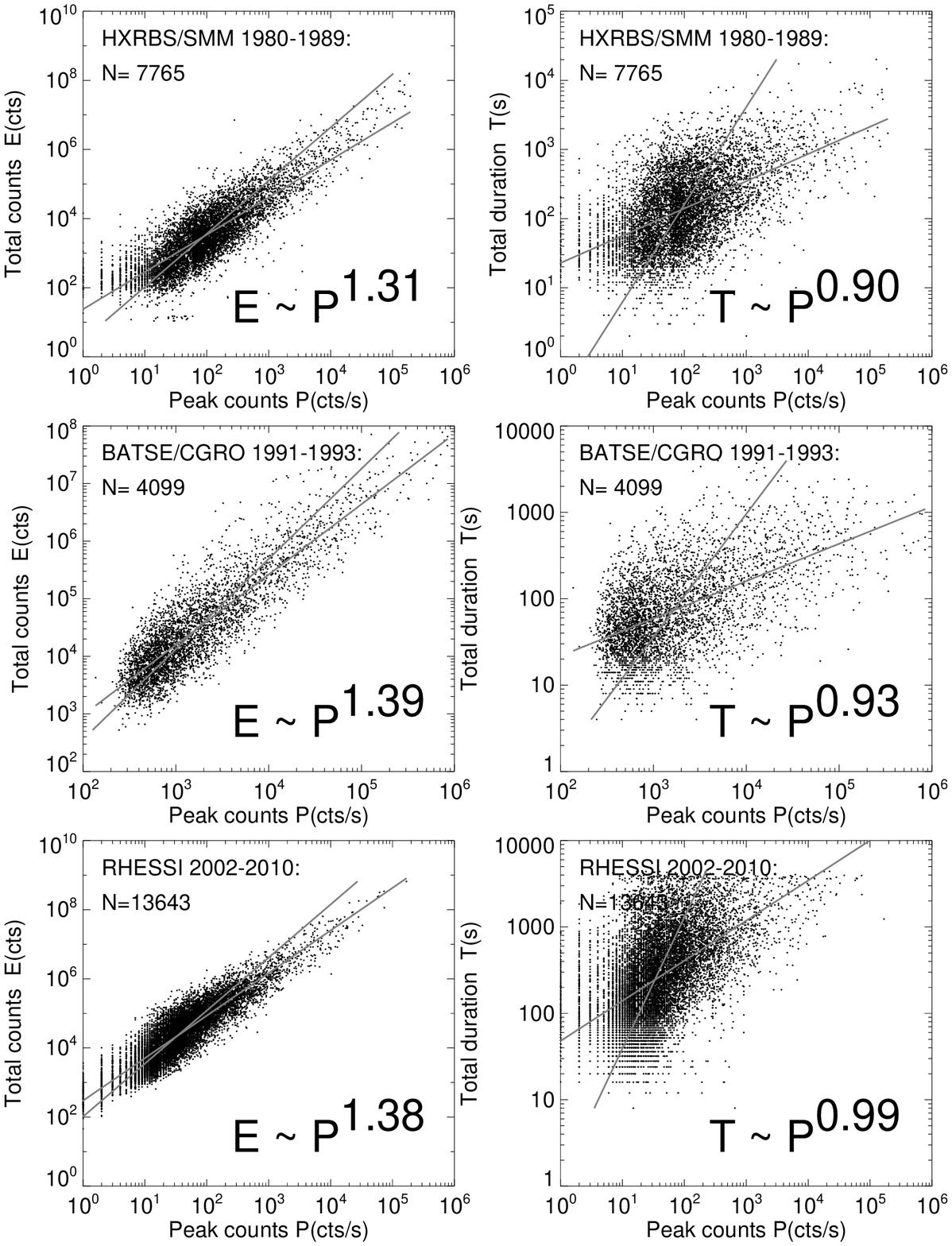}}
\caption{Scatterplots between the total counts $E(P)$ (left panels)
or flare duration $T(P)$ (right panels) versus the peak count rate $P$ 
for solar flares with HXRBS/SMM (1980 -- 1989) (top), BATSE/CGRO (1991 -- 1993) 
(middle), and RHESSI (2002 -- 2010) (bottom). Linear regression fits 
are applied for $y(x)$ and $x(y)$ (grey lines) and the listed values
correspond to the average of the two linear regression fits.}
\end{figure}

Using data from a $>25$ keV hard X-ray detector onboard the ISEE--3/ICE 
spacecraft during 24 August 1978 and 11 July 1986, Lu \etal(1993) determined 
the frequency distributions of the peak luminosity $P$ (erg s$^{-1})$, the 
energy $E$ (erg), and flare duration $T$ (s) and found that the measured 
distributions could be best fitted with a cellular automaton model that 
produced powerlaw slopes of $\alpha_P=1.86$, $\alpha_E=1.51$, and 
$\alpha_T=1.88$. The fits of the distributions included an exponential 
rollover at the upper end, which explains why they inferred
a less steep slope for durations than previously reported. 
Lee \etal(1993) analyzed
the same data and determined the correlations and frequency distribution
powerlaw slopes with special care of truncation biases and obtained similar
values for ISEE--3 ($\alpha_P=1.75$, $\alpha_E=1.62$, $\alpha_T=2.73$) as 
Crosby \etal(1993) for HXRBS. A third study was done with the same data
(Bromund \etal1995), where the energy spectrum was also calculated to
determine different energy parameters, similar to the study of 
Crosby \etal(1993), finding the following powerlaw slopes:
$\alpha_P=1.86,...,2.00$ for the peak photon flux $P_{ph}$ (photons cm$^{-2}$ s$^{-1}$),
$\alpha_P=1.92,...,2.07$ for the peak electron power $P_{e}$ (erg s$^{-1}$),
$\alpha_E=1.67,...,1.74$ for the total electron energy $E_{e}$ (erg), and 
$\alpha_T=2.40,...,2.94$ for the total duration $T$ (s), where the range
of powerlaw slopes results from the choice of the fitting range.
The flare duration $T$ was defined at a level of 1/e times the peak count rate.

From the latest solar mission with hard X-ray capabilities, the
{\sl Ramaty High-Energy Solar Spectroscopic Imager} (RHESSI) spacecraft,
frequency distributions were determined in the 12 -- 25 keV energy band
from 2002 -- 2005 (Su \etal2006), finding powerlaw slopes of 
$\alpha_P=1.80\pm0.02$ for the peak fluxes, and a broken powerlaw 
$\alpha_T$=0.9 -- 3.6 for the flare duration, similar to previous findings 
(e.g., Crosby \etal1998). Christe \etal(2008) conducted a search of
microflares and identified a total of $\approx 25,000$ events observed
with RHESSI during 2002 -- 2007 and investigated the frequency distributions
at lower energies, finding powerlaw slopes of 
$\alpha_P=1.50\pm0.03$ for 3 -- 6 keV peak count rates $P_{ph}$ (cts s$^{-1}$),
$\alpha_P=1.51\pm0.03$ for 6 -- 12 keV peak count rates, and
$\alpha_P=1.58\pm0.02$ for 12 -- 25 keV peak count rates.
Converting the peak count rates $P$ into total energy fluxes by integrating
their energy spectra, Christe \etal(2008) find an energy distribution with
a powerlaw slope of $\alpha_E=1.7\pm0.1$, with an average energy deposition
rate of $\lapprox 10^{26}$ erg s$^{-1}$. It is interesting that this
microflare statistics is fairly consistent with overall flare statistics,
even though it represents only a subset in the lowest energy range.

Flare statistics was also gathered from the {\sl Solar X-ray/Cosmic Gamma-Ray
Burst Experiment} (GRB) onboard the {\sl Ulysses} spacecraft (Tranquille
\etal2009), finding similar results for $>25$ keV events, \ie, a powerlaw
slope of $\alpha_P=1.60\pm0.04$, which steepens to $\alpha_P=1.75\pm0.08$
if the largest events with pulse pile-up are excluded.  

\medskip
We re-compiled statistics from existing flare catalogs from 
the three instruments HXRBS/SMM (1980 -- 1989), BATSE/CGRO (1991 -- 1993), 
and RHESSI (2002 -- 2010) and show summary plots of the resulting 
frequency distributions in Figures 1--3. The powerlaw slopes are obtained 
from weighted linear regression fits, using the Poisson statistics of
the number of events in each logarithmic bin (see Section 3.5).
The average powerlaw slope for peak fluxes $P$ is $\alpha_P=1.72\pm0.08$ 
(Figure 1). The corresponding frequency distributions of total counts 
or fluences are shown in Figure 2, which have an average powerlaw slope 
of $\alpha_E=1.60\pm0.14$ in the range of $E \ge 10^5$ cts. 
The distributions of flare durations are shown in Figure 3, which 
exhibit an average of $\alpha_T=1.98\pm0.35$, with a tendency of 
a rollover at the low end. Thus, our synthesized reference values are,
$$
	\begin{array}{ll}
	N(P) \propto P^{-\alpha_P} \quad &\alpha_P=1.72\pm0.08 \\
	N(E) \propto E^{-\alpha_E} \quad &\alpha_E=1.60\pm0.14 \\
	N(T) \propto T^{-\alpha_T} \quad &\alpha_T=1.98\pm0.35 \\
	\end{array}
	\eqno(1)
$$
which are compatible with most of the published values listed in Table 1.
Thus, we can consider these synthesized values as representative means,
averaged from three major missions over the last 30 years and three
solar cycles, which can serve as a reference for the overall flare
productivity of the Sun in hard X-ray wavelengths. 

\medskip
In Figure 4 we show the correlation plots between the parameters (sampled
in Figures 1 -- 3) and determine linear regression fits, which yield 
scaling laws of $E \propto P^{1.36\pm0.04}$ and $T \propto P^{0.94\pm0.05}$.
Theoretically, these correlation coefficients can also be calculated from
the powerlaw slopes of the frequency distributions. If two parameters
$x$ and $y$ have powerlaw distributions $N(x) \propto x^{-\alpha_x}$ and
$N(y) \propto y^{-\alpha_y}$, and the parameters are correlated by a
powerlaw relationship $y \propto x^{\beta}$, the powerlaw indices are
related by $\beta = (\alpha_x-1)/(\alpha_y-1)$ (e.g., Aschwanden 2010a;
Section 7.1.6). Therefore, based on the powerlaw slopes $\alpha_P$,
$\alpha_E$, and $\alpha_T$ measured in Figures 1--3, we expect the following
correlations between these three parameters,
$$
        \begin{array}{ll}
        E \propto P^{\beta_E} \quad &\beta_E={(\alpha_P-1) / (\alpha_E-1)}
                = {(1.72-1)/(1.60-1)}=1.2\pm0.2 \\
        T \propto P^{\beta_T} \quad &\beta_T={(\alpha_P-1) / (\alpha_T-1)}
        	= {(1.72-1)/(1.98-1)}= 0.8\pm0.2\\
        \end{array}
        \eqno(2)
$$
The correlation plots shown in Figure 8, where the linear regression fit
is performed for $y(x)$ with $x$ as independent variable, as well as
for $x(y)$ with $y$ as independent variable, respectively. 
The averaged powerlaw slopes
(of both linear regression fits and all three datasets)
yield values of $E \propto P^{1.36\pm0.04}$ and $T \propto P^{0.94\pm0.05}$,
which are fully consistent with the values derived from the powerlaw
slopes of the frequency distributions (Eq.~2), i.e., 
$E \propto P^{1.2\pm0.2}$ and $T \propto P^{0.8\pm0.2}$.

\section{Uncertainties and Errors of Powerlaw Slopes}

The determination of powerlaw slopes and their uncertainties are subject
to methodical and data selection effects. Given the variation of observed 
powerlaw slopes as listed in Table 1, it might be useful to briefly discuss 
some important selection effects and statistical uncertainties. 

\subsection{Energy or Wavelength Bias}

In Table 1 we compare mostly flare data in the hard X-ray regime
with energies of $>20$ keV or $>25$ keV, which all have similar powerlaw 
slopes of $\alpha_P \approx 1.6 - 1.8$.
Even flares detected at lower energies $>8$ keV 
(Lin et al.~2001) or $>12$ keV (Su et al.~2006; Christe et al.~2008)
have similar powerlaw slopes. On the other end, flares detected at 
high energies of $>100$ keV (Perez Enriquez and Miroshnichenko 1999) have 
also similar powerlaw slopes. However, the same subset of flares observed
at 0.075 -- 124 MeV, from bremsstrahlung at 300 -- 850 keV,
from the 511 keV annihilation line fluence,
from the 2.223 MeV neutron line fluence, and
from the 1 -- 10 MeV gamma-ray line fluence,  
have significantly flatter powerlaw slopes ($\alpha_P \approx 1.3-1.4$).
This can be explained by the flatter energy spectra of large flares,
which causes relatively higher fluxes at higher energies, and thus
flatter occurrence frequency distributions. Dennis (1985) has shown
a systematic tendency of the hard X-ray spectrum as a function of the
hard X-ray count rate, being generally flatter for large flares with
high count rates, which predicts also flatter powerlaw slopes
of the occurrence frequency distributions at higher energies.  
Similar wavelength bias effects occur also for flare fluxes detected
in soft X-rays and EUV wavelengths, compared with hard X-ray fluxes,
which depend on the physical emission mechanism. Therefore we confine
our comparisons in Table 1 only to hard X-ray energies in the range of
$>8$ keV to $>100$ keV here. 

\subsection{Effects of Data Gaps and Spacecraft Orbits}

Data gaps due to spacecraft night or South Atlantic Anomaly (SAA)
passages can also cause systematic biases, because they
tend to shorten flare durations (due to interruption) and underestiamte
the total flare counts (due to incomplete sampling). These effects can in
principle be simulated and their systematic effects on frequency 
distributions can be quantified this way. Numerical simulations
of such selection and instrumental effects 
are also discussed in the context of waiting time distributions
(\eg, Aschwanden and McTiernan 2010). The distribution of peak fluxes
is less affected by data gaps, as long as the peak of a flare is observed.
Visual screening of flare light curves was carried out for events detected
with HXRBS/SMM, so that missed flare peaks could be identified in the
flare catalog (Dennis, private communication). For RHESSI flare catalogs,
however, no correction for spacecraft night datagaps has been applied to
flare peak fluxes and fluences so far, and long-duration flares are
counted multiple times for each spacecraft orbit (McTiernan, private
communication). 

\begin{figure}
\centerline{\includegraphics[width=1.0\textwidth]{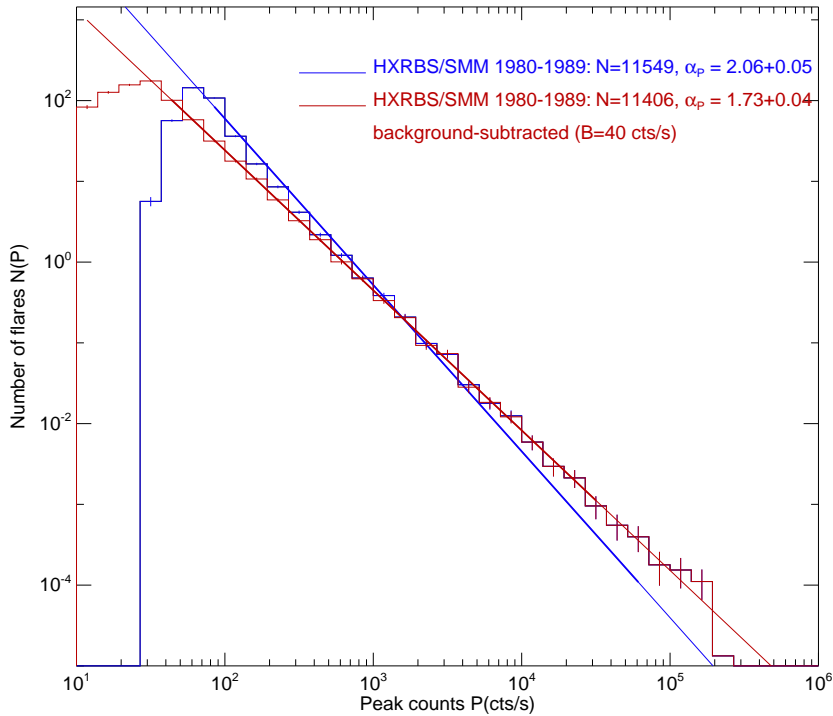}}
\caption{Occurrence frequency distributions of hard X-ray peak count rates
$P$ (cts/s) observed with HXRBS/SMM (1980 -- 1989), without background
subtraction ($\alpha_P=2.06\pm0.05$; blue), and with subtraction of
an average background of $B=40$ cts s$^{-1}$ ($\alpha_P=1.73\pm0.01$; red).}
\end{figure}

\subsection{Pre-Event Background Subtraction}

Most flare catalogs list the peak counts of flares without preflare
background subtraction, which has little impact on large flares, but
leads to a strong overestimate of fluxes for the samllest flares,
and thus systematically overestimates the powerlaw slope. We demonstrate
this for the dataset of HXRBS/SMM flares. Figure 5 shows two (weighted)
powerlaw fits, a steeper slope of $\alpha_P=2.06\pm0.05$ for the uncorrected
flare peak counts, and a flatter slope of $\alpha_P=1.73\pm 0.04$
after subtraction of an estimated average background of 40 cts s$^{-1}$.
This correction was also applied in the original analysis in Crosby
et al.~(1993) and we use it here too. Note that the background-uncorrected 
distribution shows also a steepening of the powerlaw slope at the lower end.

For the RHESSI flare catalog, the preflare background
is estimated to be $\approx 10-20$ cts s$^{-1}$ per detector, the low
limit applying to low spacecraft latitude, the high limit for high
latitutes, with a mean of $\approx 12$ cts s$^{-1}$ per detector
(Jim McTiennan, private communication).

The subtraction of an average preflare background value is only a
first-order correction, which may cause even events with negative
peak fluxes when the background is time-varying. The optimum method 
would be to measure the preflare background for each event separately 
and to subtract it from the peak fluxes, as well as from the fluences 
before time integration. 

\subsection{Instrument Sensitivity}

The instrumental sensitivity usually translates into a fixed flux threshold,
which causes a sharp lower cutoff for peak flux distributions 
(Figures 1 and 5), 
but a gradual roll-over for total fluence and duration of events 
(Figure 2 and 3). The roll-over limits the scale-free range of the
distribution over which a powerlaw can be fitted, and moreover produces
a systematic bias for flatter powerlaw slopes when part of the roll-over
is included in the powerlaw fit. If powerlaw fits are weighted by the
number of events per bin in a linear regression fit to a histogram,
the lowest bins will have the hightest weight, and thus a gradual
roll-over at the lower end will produce a too flat powerlaw slope,
when compared with the fit in the upper bins of the distribution.
We determine the lower bound of the powerlaw fit range by the
$\chi^2$-criterion, which reliably detects a roll-over by an excessive
deviation from a powerlaw fit in terms of the expected standard deviation.  

\subsection{Linear Regression Fit Methods}

A common method to determine powerlaw slopes is a linear regression fit
to a histogram in a $log(N)-log(S)$ diagram, which can be done in two ways.
If the frequency distribution closely follows a powerlaw 
function, a weighted linear regression fit can be performed in log-log
space, using the number of events per bin as relative weights. However, 
if the frequency distribution exhibits significant deviations from a 
straight powerlaw
(\eg, Figures 2 and 3), a weighted powerlaw fit can only be performed
in a powerlaw-like subinterval. Often an unweighted fit (with equal weight
in log-log space) is carried out in such cases in order to obtain an
approximate powerlaw slope. For the results given in literature (as
compiled in Table 1), it is not always clear whether the authors performed
a weighted or unweighted linear regression fit, which may significantly
deviate from each other depending on the fitted interval. In this study here,
we perform only weighted fits in well-defined intervals as described below.

In a log-log histogram, the bins $\Delta x_i$ are chosed to be equidistant 
on a log-scale, but they are exponentially increasing on a linear scale. 
If $N(x) \approx x^{-\alpha}$ is the fitted powerlaw function, the number 
of events $N_i$ per bin is,
$$
	N_i = \int_{x_i}^{x_i+\Delta x_i} N(x) dx
	    \approx N(x_i) \Delta x_i \ .
	\eqno(3)
$$
The statistical uncertainty $\sigma_i^{log}$ in bin $\Delta x_i$ due to
Poisson statistics is then on a log-scale,
$$
	\sigma_i^{log} = \log(N_i + \sqrt{N_i}) - \log(N_i) \ ,
	\eqno(4)
$$
and the statistical weighting factor $w_i$ for a linear regression fit 
that minimizes the squared standard deviations $\sigma_i^{log}$ in
log-log space is then,
$$
	w_i = {1 \over [\sigma_i^{log}]^2} \ .
	\eqno(5)
$$
The linear regression fit in log-log space finds then the following
best-fit powerlaw function,
$$
	log(N_i^{fit}) = N_0 + \alpha \log(x_i) \ , 
	\eqno(6)
$$
A criterion for the goodness of a fit is the $\chi^2$-test, expressed
with the so-called {\sl reduced $\chi^2$-value}, which expresses the
mean deviation of the data points from the fitted function in units
of expected standard deviations,
$$
	\chi_{red} = \left[ {1 \over (n-n_{free})} \sum_{i=1}^n 
		{(N_i - N_i^{fit})^2 \over (\sigma_i^{log})^2}
		\right]^{1/2} \ ,
	\eqno(7)
$$
where $n$ is the number of datapoints (or histogram bins here) and
$n_{free}$ is the number of free parameters of the fitted function,
which is $n_{free}=2$ for a linear regression fit (with free parameters
$N_0$ and $\alpha$ in Eq.~6). However, since the upper part of the
histogram in log-log distributions generally include bins with zero 
or very few events, the validity of Poisson statistics $\sqrt{N_i}$
is hampered for $N_i \lapprox 10$ and has to be replaced by
C-statistics (Cash 1979), which has the following expression 
corresponding to the reduced $\chi^2$-criterion,
$$
	\chi_{cash} = \left[ {2 \over (n-n_{free})} \sum_{i=1}^n 
		N_i \ln(N_i/N_i^{fit}) - (N_i - N_i^{fit}) \right]^{1/2} \ ,
	\eqno(8)
$$
where $N_i^{fit}$ is the theoretical expectation value based on the
best-fit powerlaw function and has to be positive for all fitted bins.
For an acceptable powerlaw fit, a value of $\chi_{Cash} \approx 1.0$
is expected. The linear regression fit yields then a powerlaw slope
$\alpha$ and the probable uncertainty $\sigma_\alpha$ due to
Poisson statistics, or C-statistics, respectively. 

\begin{figure}
\centerline{\includegraphics[width=1.0\textwidth]{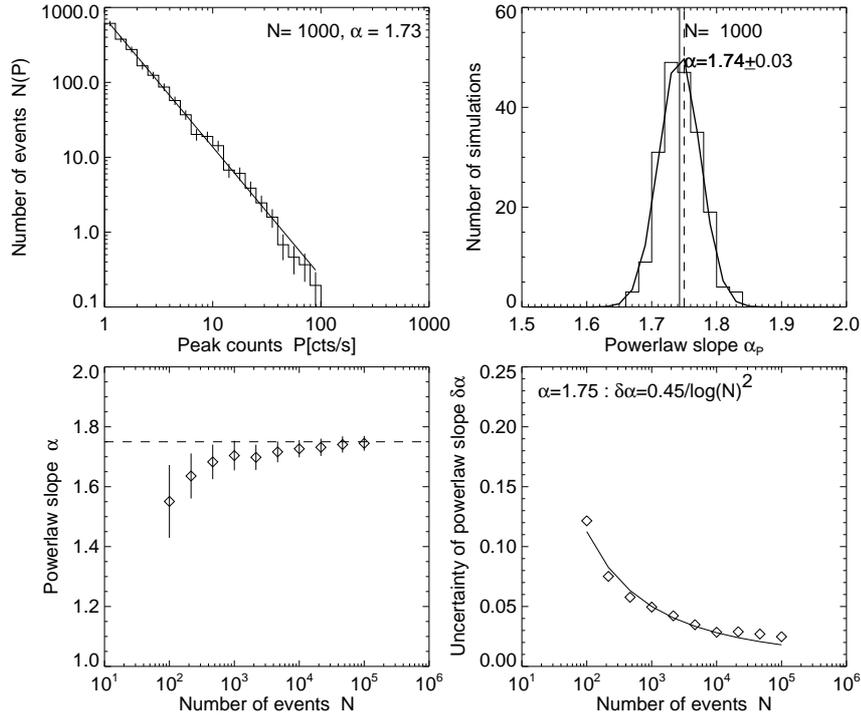}}
\caption{Monte-Carlo simulations of powerlaw distributions 
are carried out for a theoretical powerlaw slope of 
$\alpha =1.75$ with different random representations according
to Equation 11. A single distribution based on $N=1000$ events is
shown with powerlaw fit $\alpha=1.73$ (top left panel). The
Gaussian distribution of powerlaw slopes obtained from $N_{run}=
1000$ runs with different random representations yields a mean of
$\alpha=1.74\pm0.03$ (top right). These Monte-Carlo simulations
are repeated for various number of events $N=10^2,...,10^5$, which
show a convergence to the mean value and standard deviations
towards the theoretical value $\alpha=1.75$ (bottom left).
The scaling of the uncertaingy $\delta \alpha(N)$ as a function of
the number of events $N$ is also shown (bottom right).}
\end{figure}

\subsection{Monte-Carlo Simulations of Powerlaw Uncertainty}

An alternative method to infer the uncertainty of powerlaw slopes
is a Monte-Carlo simulation of generating a powerlaw distribution.
Mathematically, a desired distribution function $f(x)$ can be
generated from uniformly distributed random values $\rho$ in the
range of $[0,1]$ by using the inverse function $F^{-1}(x)$ of the integral
function $F(x)$ of $f(x)$. Here we wish to generate a
frequency distribution of energies $E$ in the form of a powerlaw 
function $p(E)$ with slope $\alpha$ (Aschwanden 2010a, Section 7.1.5),
$$
        p(E) = (\alpha - 1) E^{-\alpha} \ ,
        \eqno(9)
$$
which fulfills the normalization $\int_1^{\infty} p(E) dE=1$.
The total probability $\rho(E)$ in the range $[0,E]$ is
then the integral function of $p(E)$ (Eq.~9),
$$
        \rho(E) = \int_0^E p(\epsilon) d\epsilon
             = \int_0^E (\alpha - 1) \epsilon^{-\alpha} d\epsilon
             = \left[ 1 - E^{(1-\alpha)} \right] \ .
        \eqno(10)
$$
The inverse function $E(\rho)$ of $\rho(E)$ (Eq.~10) is
$$
        E(\rho) = \left[ 1 - \rho \right]^{1/(1-\alpha)} \ .
        \eqno(11)
$$
In Figure 6 (top left) we use a random generator that produces 1000 values
$\rho_i$ uniformly distributed in the range of $[0,1]$, choose a powerlaw 
index of $\alpha=1.75$, and use the transform Eq.~(11) to generate values
$E_i = [1 - \rho_i]^{1/(1-\alpha)}$ and sample the frequency distribution 
of the 1000 values $E_i$, which produces a powerlaw function 
$p(E) =(\alpha-1) E^{-\alpha}$ as defined in Eq.~(9), with a best-fit
slope of $\alpha=1.73$.

We repeat the generation of powerlaw histograms with $N=1000$ different
random representations $\rho_i$ and find for the powerlaw slopes
a Gaussian distribution with a mean and standard deviation of
$\alpha=1.74\pm0.03$ (Figure 6, top right). The spread of 
$\alpha$-values is expected to depend on the number $N$ of events 
in a distribution. In order to quantify this dependence we repeat 
the same exercise for different numbers of events in the range of 
$N=10^2, ..., 10^5$, which is shown in Figure 6 (bottom left). The
standard deviations of the powerlaw slope distributions is found
to depend on the number of events (per distribution) as 
$\delta \alpha \approx {0.45 / \log(N)^2 }$ (Figure 6, bottom right). 
Repeating the same Monte-Carlo simulations for a range of powerlaw slopes
($\alpha = 1.5, ..., 2.0$) we find the following general dependence, 
$$
	\delta \alpha(\alpha, N) \approx (\alpha-0.75) 
			{0.45 \over \log(N)^2 } \ .
	\eqno(12)
$$
This uncertainty of the powerlaw slope obtained from Monte-Carlo
simulations is typically a factor of $\approx 1.5\pm0.4$ larger than 
the formal uncertainty obtained from linear regression fits (according
to the statistics given in Table 2). Both methods are based on the
Poisson statistics of random numbers and are expected to lead to 
commensurate uncertainties of powerlaw slopes. However, while the
uncertainty in a linear regression fit is based on a single dataset,
the uncertainty from Monte-Carlo simulations includes in addition the
scatter of many random representations and thus yields a slightly
higher uncertainty. Thus, we consider the latter method as a more reliable 
measure or uncertainties when comparing the best-fit powerlaw slopes
among two different datasets.

\begin{figure}
\centerline{\includegraphics[width=1.0\textwidth]{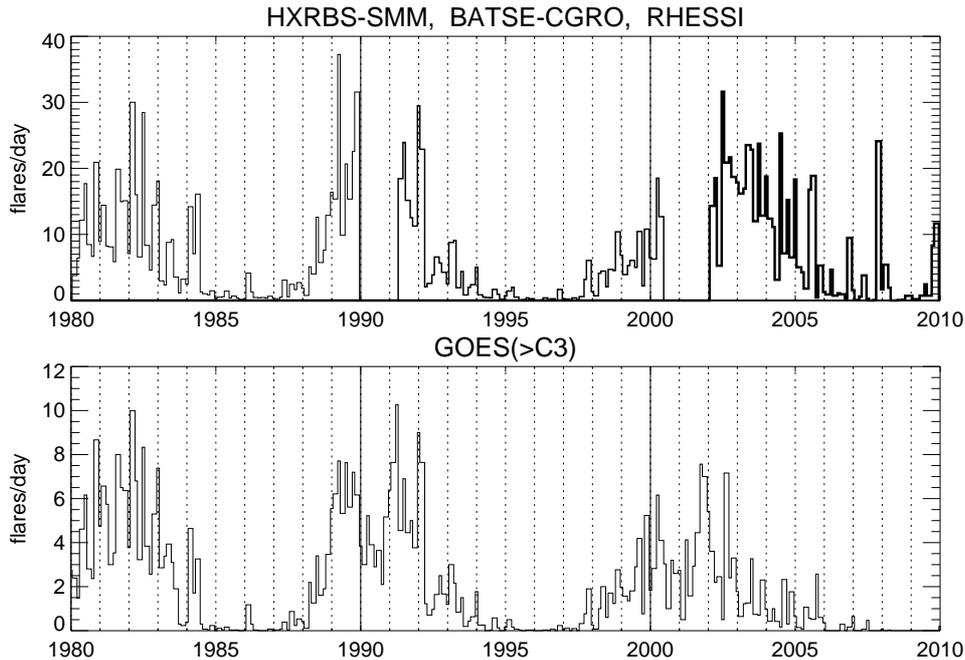}}
\caption{Top: Monthly averages of solar flare rates observed during the
last three solar cycles in hard X-rays with HXRBS/SMM (1980 -- 1989), 
BATSE/CGRO (1991 -- 2000), and RHESSI (2002 -- 2010). 
Bottom: Monthly averages of solar flare rate observed in soft X-rays
with GOES, including events above the C3-class level.}
\end{figure}

\section{Solar Flare Statistics versus Solar Cycle}

We investigate now whether the statistical distributions of the most
reliably determined flare parameter, namely the peak flux $P$,
exhibits significant changes during the solar cycles. 
From the time profiles of the monthly flare rates shown for
solar cycles 21, 22, and 23 (Figure 5), we identify approximately
the following time intervals for the solar minima: 1985 -- 1987, 
1995 -- 1997, and 2007 -- 2010. 

For the first solar minimum we have coverage by HXRBS/SMM data.
We break down the dataset into 2-year intervals and show the frequency 
distributions of the peak flux separately for 1980-1981, 1982-1983, 
1984-1985, 1986-1987, and 1988-1989 in Figure 8 (left column). 
We perform weighted linear regression fits in the lower half
of the distributions, bracketed by the values $P_1 < P < P_2$ (dotted
lines in Figure 8), where $P_1$ corresponds to the next bin above the 
maximum of the frequency distribution, and $P_2$ is chosen to be higher
by a constant factor of 20. The number of events $n$, the lower and upper bounds
$P_1$ and $P_2$, the powerlaw fits $\alpha_P$ with the formal fit
uncertainty $\sigma_\alpha$, the uncertainty $\delta \alpha_P$ obtained
from the Monte-Carlo simulations (Equation 12), and the reduced
$\chi^2$-value according to C-statistics ($\chi_{Cash}$) are all
listed in Table 2 and shown in Figure 8. We are doing the same procedure
for the BATSE/CGRO dataset by breaking it down into the two intervals
of 1991 and 1992--1993, which covers part of Solar Cycle 22. 
In addition, we perform the same procedure for RHESSI data by
breaking down the dataset into the following 2-year intervals:
2002--2003, 2004-2005, 2006-2007, 2008-2009, and a 1-year interval
of 2010. Thus, we determined the powerlaw slope in the lower half
of the frequency distribution for a number of 12 time intervals with
a length of 1-2 years during the three different solar cycles.

\begin{table}
\begin{center}
\footnotesize	
\caption{Frequency distributions measured from solar flares in hard X-rays
during the solar cycle.}
\begin{tabular}{lrrrrrr}
\hline
Instrument &Years & Number & Lower & Upper & Powerlaw                  & Goodness \\
       &          & of     & fit   & fit   & slope of                  & of fit   \\
       &          & events & bound & bound & peak counts               &          \\
       &          & $n$    & $P_1$ & $P_2$ & $\alpha_P$ ($\delta\alpha_P$) &$\chi_{Cash}$\\
\hline
HXRBS  &1980-1981 &   4460 &    43 &   848 &  1.77$\pm$0.02 ($\pm$0.05) &  1.43 \\
HXRBS  &1982-1983 &   2642 &    43 &   848 &  1.77$\pm$0.03 ($\pm$0.05) &  0.66 \\
HXRBS  &1984-1985 &    581 &    43 &   848 &  1.91$\pm$0.06 ($\pm$0.08) &  1.16 \\
HXRBS  &1986-1987 &    437 &    43 &   848 &  1.75$\pm$0.08 ($\pm$0.07) &  1.03 \\
HXRBS  &1988-1989 &   3286 &    31 &   439 &  1.60$\pm$0.03 ($\pm$0.04) &  1.34 \\
BATSE  &1991-1991 &   2389 &   610 & 11787 &  1.61$\pm$0.03 ($\pm$0.05) &  0.56 \\
BATSE  &1992-1993 &   1720 &   610 & 11787 &  1.69$\pm$0.03 ($\pm$0.05) &  0.90 \\
RHESSI &2002-2003 &   6678 &    31 &   439 &  1.59$\pm$0.02 ($\pm$0.04) &  1.69 \\
RHESSI &2004-2005 &   3930 &    31 &   439 &  1.66$\pm$0.03 ($\pm$0.04) &  1.78 \\
RHESSI &2006-2007 &    958 &    31 &   439 &  1.83$\pm$0.06 ($\pm$0.07) &  1.31 \\
RHESSI &2008-2009 &    581 &    31 &   610 &  1.83$\pm$0.05 ($\pm$0.07] &  0.65 \\
RHESSI &2010-2010 &   1496 &    43 &   848 &  1.89$\pm$0.04 ($\pm$0.06] &  0.79 \\
\hline
\end{tabular}
\end{center}
\end{table}

\begin{figure}
\centerline{\includegraphics[width=1.0\textwidth]{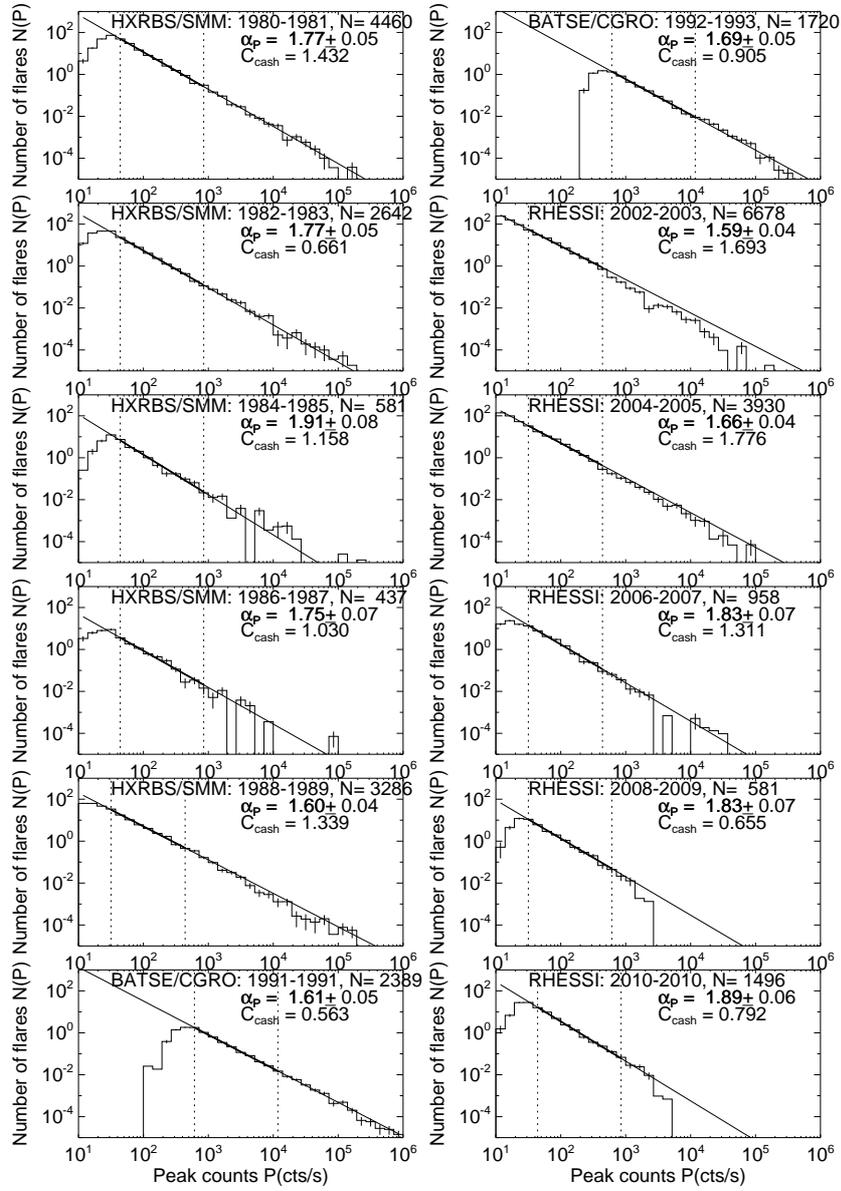}}
\caption{Occurrence frequency distributions of 
hard X-ray peak count rates $P$ (cts/s) observed with HXRBS/SMM,
BATSE/CGRO, and RHESSI, broken down into 2-year intervals.
The powerlaw fits were determined with weighted linear regression
fits applied to the lower half of the sampled distributions ,
within a range ratio of $P_2/P_1=20$ (dotted vertical lines 
and thick linestyle for best-fit powerlaw functions). 
The best-fit values are also listed in Table 2 and plotted as a
function of time in Figure 9.}
\end{figure}

In order to investigate the time dependence of the powerlaw slopes
of peak flux distributions we plot the values $\alpha_P \pm \delta \alpha_P$
in the middle panel of Figure 9, which exhibits a sinusoidal variation
during the solar cycles. For comparison, we plot also the published
values of datasets with a timespan less than a half solar cycle
($\le 6$ years) and a minimum of $n>100$ events, which includes
datasets 1, 3--8, and 12--14 from Table 1 (Figure 9, top panel). 
The datasets labeled
with 17--19 in the middle panel correspond to HXRBS, BATSE, and RHESSI
as listed in Table 2 and analyzed in Figures 1--4 and 8. 

We are now characterizing the time variation of the powerlaw slope
$\alpha_P(t)$ as shown in Figure 9. First we test the hypothesis
whether the data are consistent with a constant value, using simply
the $\chi^2$-criterion (Equation 7) for the expectation value of
a constant mean value (horizontal lines in Figure 9). 
For the published values we find a mean of
$\alpha_P=1.72\pm0.08$ and a $\chi_{red}=2.01$ (Figure 9 top panel), 
and similarly for our re-analyzed powerlaw slopes with weighted
linear regression fits in the lower half of the frequency distributions, 
\ie $\alpha_P=1.74\pm0.11$ and $\chi_{red}=2.7$ (Figure 9 middle panel). 
Based on this unacceptable goodness-of-fit value $\chi_{red} \gapprox 2$ 
we can reject the hypothesis that the powerlaw index is constant during 
the solar cycles.

Next we test the hypothesis whether the powerlaw slope $\alpha_P(t)$
varies with a sinusoidal variation,
$$
	\alpha_P(t)=\alpha_0+\Delta \alpha \cos{[2\pi (t-t_0)/T_{cycle}]} \ .
	\eqno(13)
$$
For the published values, using the uncertainties that are quoted
in the publications, we find no acceptable fit, based on a goodness-of-fit
value of $\chi_{red}=2.5$ (Figure 9, top panel). The discrepancy comes
mostly from RHESSI data, where two quite different values were measured
during the same time interval, namely $\alpha_P=1.80\pm0.02$ (dataset 13;
Su \etal 2006) versus $\alpha_P=1.58\pm0.02$ (dataset 14; Christe \etal 2008). 

\begin{figure}
\centerline{\includegraphics[width=1.0\textwidth]{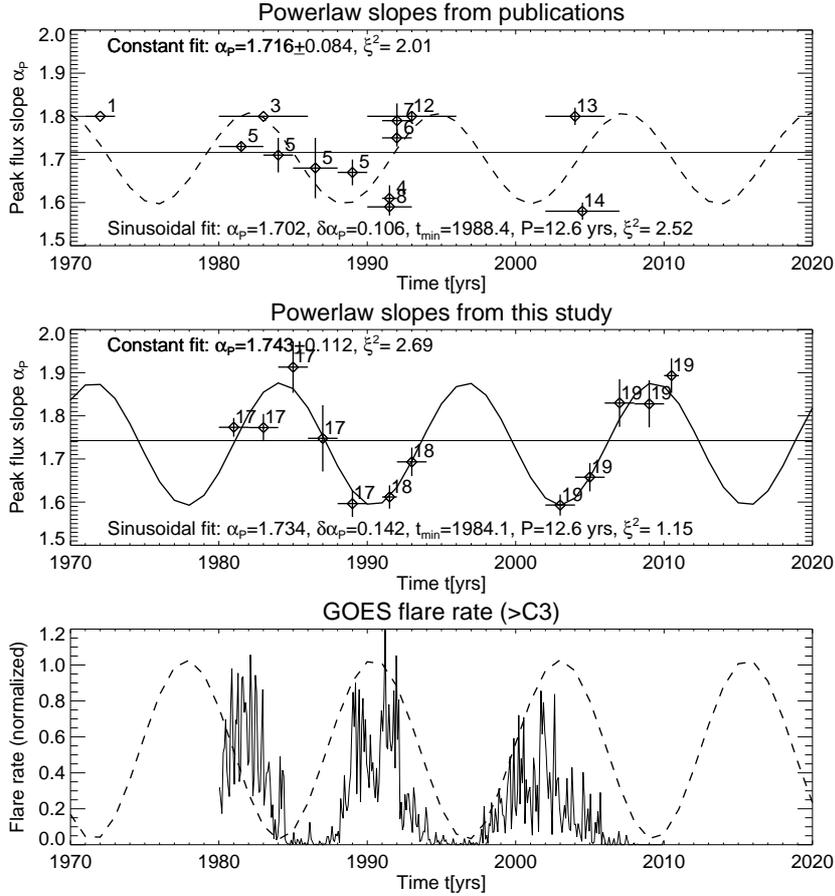}}
\caption{The powerlaw slope $\alpha_P$ of the peak flux $P$ as
a function of time during the last three solar cycles. The diamond
symbols and numbers indicate the mean values $\alpha_P$ and the
corresponding reference number quoted in Table 1. The horizontal
lines indicate the time intervals of observation and the vertical
bars indicate the cited uncertanties. The functional dependence of
the powerlaw slope $\alpha_P(t)$ on the solar cycle is tested with 
a fit of a constant value (horizontal lines) and with a sinusoidal
curve (dashed curve in top panel, and thick curve in middle panel).
The bottom panel shows the GOES flare rate for comparison (see also
Figure 7).}
\end{figure}

For the powerlaw fits obtained in this study, however, we find an
acceptable fit with $\chi_{red}=1.15$ and best-fit parameters of:
mean powerlaw slope $\alpha_P=1.73$, amplitude of variation
$\Delta \alpha =0.14$, solar cycle period $T_{cycle}=12.6$ yrs,
and a cycle minimum time $t_0=1984.1$ (Figure 9, middle).
This is a novel result that shows for the first time the solar cycle
dependence of the powerlaw slope of flare hard X-ray count rate 
distributions. For comparison we plot the monthly GOES flare rate
(for $>C3$ class) in the bottom panel of Figure 9 (also shown in Figure 7).
The variation of the powerlaw slope is clearly anti-correlated to
the flare rate or sunspot number, bein steepest during solar minima
and flattest during solar maxima. The best-fit solar cycle period
is $T_{cycle}=12.6$ years applies to the mean behavior during the fitted
period of 1980-2010, but is heavily weighted by the longest last 
Cycle 23. The actual lengths of solar cycles published by NOAA are
10.3 years for Cycle 21 (1976-1986), 
9.7 years for Cycle 22 (1986-1996), and
12.6 years for Cycle 23 (1996-2006). 

\section{Discussion and Conclusions}

Frequency distributions of solar-flare parameters have been reported 
over the last three solar cycles from different instruments and different
phases of the solar cycle, covering a scattered range of values (Table 1).
Powerlaw slopes were reported in a range of
$\alpha_P \approx 1.6 - 2.0$ for peak fluxes (counts),
$\alpha_E \approx 1.4 - 1.7$ for total fluxes (fluences), and
$\alpha_T \approx 1.9 - 2.8$ for total durations, if we restrict
to the hard X-ray range of $\approx 10-100$ keV.
Averaging the powerlaw slopes from the largest
datasets spanning entire solar cycles (using data from 
HXRBS/SMM, BATSE/CGRO, RHESSI), we find mean values of 
$\alpha_P = 1.72\pm0.08$ for peak fluxes (Figure 1),
$\alpha_E = 1.60\pm0.14$ for fluences (Figure 2), and
$\alpha_T = 1.98\pm0.35$ for flare durations (Figure 3).  
Investigating variations of the flare statistics on time scales of
1-2 years, however, we do find evidence for a significant variation of the 
powerlaw slope of the frequency distributions during the solar cycle.
We find that the the powerlaw slope varies sinusoidally with an amplitude 
of $\pm 8\%$ during the solar cycle, in anti-correlation to the sunspot number,
being flattest during the maximum of the solar cycle.

Why is our obtained result of the anti-correlation of the powerlaw slope
of flare count rates with the solar cycle not evident 
from previously published values?
First of all, the datasets have been subject to inconsistent analysis
methods by different authors, differing in the type of linear regression fits
(weighted vs. unweighted), the fitted range $[P_1, P_2]$, 
selection effects of events (flare catalogs vs. automated microflare
search algorithms), background treatment (with or without background
subtraction), as discussed in Section 3. In order to identify some
systematic effects we varied the fitting range $[P_1, P_2]$ between
one to three decades and found that the sinusoidal variation shows
up clearest for a factor of $P_2/P_1=20$, which entails the lower
half of frequency distributions, typically GOES C-class to M-class
flares, but excludes the large GOES X-class flares that make up
the upper half of the frequency distributions. Obviously, X-class
flares tend to appear more irregularly during the solar cycles,
while the more frequently C-class and M-class flares vary in a better 
synchronized way with the solar cycle. The fact that the Sun produces
flatter powerlaw slopes during periods of high flare activity
indicates that the magnetic complexity is higher during activity
periods and produces over-proportionally energetic flares during
these periods. The effect is analog to the soft-hard-soft evolution
of energy spectra known during single flares.  

Marginal variations of the powerlaw slope with flare activity
have been noted before. Crosby \etal (1993; Fig.~12 therein) 
found variations of the powerlaw slope during the solar cycle
in excess of the statistical uncertainties of the linear regression
fits, but no clear functional dependence on the solar cycle was found.
Biesecker (1994) noticed slight differences of the powerlaw slope
during low activity $(\alpha_P=1.71\pm0.04$) and high activity periods
($\alpha_P=1.68\pm0.02$), with the powerlaw slope usually flatter for
high-activity periods, which is analog to our finding of a flatter
slope during solar cycle maxima. 
Bai (1993) devised a special maximum likelihood method to determine
the powerlaw slope of flare distributions from HXRBS/SMM and found
some variation of the powerlaw slope correlated with a 154-day
periodicity of flare rates. He found that the size distrtibutions are 
steeper during the maximum years of Solar Cycle 21 (1980 and 1981)
than in the declining phase (1982-1984). The cycle effect is opposite 
to our findings, but it depends sensitively on the threshold or
fitted range. However, the three studies (Bai 1993; Biesecker 1994;
and this study) agree in the result that a flatter powerlaw slope
is correlated with a higher flare activity, either on periods of
recurrent flare activity (153.8 days) or solar cycles (12.6 years).

What does our novel result suggest for the physical interpretation?
A flatter frequency distribution implies an overproportional amount
of larger events, compared with an averaged distribution. This means
that the physical conditions are different during the solar cycle.
One possible explanation could be that the magnetic complexity
increases during the solar maximum, which produces more stressing
of magnetic fields and larger energy releases. The magnetic fields
during the maximum of the solar cycle are dominated by the toroidal
component of the solar dynamo, which entails higher magnetic stresses.
The magnetic field during the minimum of the solar cycle is simpler
and is dominated by the dipolar poloidal field that spans from the
north to the south pole. It appears that the relative amount of
energy released in flares is modulated by the magnetic field
complexity of the solar dynamo. Theoretical models that reproduce
the statistical distributions of solar flares in terms of the concept
of self-organized criticality are discussed in Paper II
(Aschwanden 2010b).

\acknowledgements 
We thank the referee Brian Dennis and James McTiernan for constructive 
and helpful comments that substantially improved this study. 
This work is partially supported by NASA contract
NAS5--98033 of the RHESSI mission through the University of California,
Berkeley (subcontract SA2241--26308PG) and NASA grant NNX08AJ18G.
We acknowledge access to solar mission data and flare catalogs from the
{\sl Solar Data Analysis Center} (SDAC) at the NASA Goddard Space Flight
Center (GSFC).

\section*{References} %%% REFERENCES 

\def\ref#1{\par\noindent\hangindent1cm {#1}}

\small
\ref Aschwanden, M.J., Dennis, B.R., Benz, A.O. 1998,
        Logistic avalanche processes, elementary time structures,
        and frequency distributions of flares,
        {\it Astrophys. J.} {\bf 497}, 972-993.
\ref Aschwanden, M.J. and McTiernan, J.M., 2010,
 	Reconciliation of waiting time statistics of solar flares 
	observed in hard X-rays,
 	{\it Astrophys. J.} {\bf 717}, 683-692.
\ref Aschwanden, M.J., 2010a,
	{\sl Self-Organized Criticality in Astrophysics.
	The Statistics of Nonlinear Processes in the Universe},
	Springer-Praxis: Heidelberg, New York (in press). 
\ref Aschwanden, M.J., 2010b,
	The State of Self-Organized Criticality of the Sun 
	During the Last Three Solar Cycles. II. Theoretical Modeling,
	Solar Physics, (this volume), subm. (Paper II).
\ref Bai, T. 1993,
	Variability of the occurrence frequency of solar flares as 
	a function of peak hard X-ray rate,
	{\it Astrophys. J.} {\bf 404}, 805-809.
\ref Biesecker, D.A., Ryan, J.M., Fishman, G.J. 1993,
        A search for small solar flares with BATSE,
        {\it Lecture Notes in Physics} {\bf 432}, 225-230. 
\ref Biesecker, D.A., Ryan, J.M., and Fishman, G.J. 1994,
        Observations of small small solar flares with BATSE,
        in {\sl High-Energy Solar Phenomena - A New Era of Spacecraft
        Measurements}, J.M. Ryan and W.T.Vestrand (eds.),
        American Inst. Physics: New York, 183-186. 
\ref Biesecker, D.A., 1994,
	{\sl On the occurrence of solar flares observed with the
	Burst and Transient Source Experiment},
	PhD Thesis, Universithy of New Hampshire. 
\ref Bromund, K.R., McTiernan, J.M., Kane, S.R. 1995,
        Statistical studies of ISEE3/ ICE observations of impulsive
        hard X-ray solar flares, {\it Astrophys. J.}, {\bf 455}, 733-745.
\ref Cash, W. 1979,
	{\sl Parameter estimation in astronomy through application of the
	likelihood ratio}, {\it Astrophys. J.} {\bf 228}, 939-947.
\ref Christe, S., Hannah, I.G., Krucker, S., McTiernan, J., Lin, R.P.
        2008, RHESSI microflare statistics. I. Flare-finding and
        frequency distributions, {\it Astrophys. J.} {\bf 677}, 1385-1394. 
\ref Crosby, N.B., Aschwanden, M.J., Dennis, B.R. 1993,
        Frequency distributions and correlations of solar X-ray
        flare parameters, {\it Solar Phys.} {\bf 143}, 275-299.
\ref Crosby, N.B. 1996,
        {\sl Contribution \`a l'Etude des Ph\'enom\`enes Eruptifs
        du Soleil en Rayons Z \`a partir des Observations de l'Exp\'erience
        WATCH sur le Satellite GRANAT},
        PhD Thesis, University Paris VII, Meudon, Paris. 
\ref Crosby, N.B., Vilmer, N., Lund, N., and Sunyaev, R. 1998,
        Deka-keV X-ray observations of solar bursts with WATCH/GRANAT:
        frequency distributions of burst parameters,
        {\it Astrophys. J.} {\bf 334}, 299-313. 
\ref Datlowe, D.W., Elcan, M.J., Hudson, H.S. 1974,
        OSO-7 observations of solar X-rays in the energy range
        10-100 keV, {\it Solar Phys.} {\bf 39}, 155-174. 
\ref Dennis, B.R. 1985,
        Solar hard X-ray bursts, {\it Solar Phys.} {\bf 100}, 465-490.
\ref Lee, T.T., Petrosian, V., McTiernan, J.M. 1993,
        The distribution of flare parameters and implications
        for coronal heating, {\it Astrophys. J.} {\bf 412}, 401-409. 
\ref Lin, R.P., Schwartz, R.A., Kane, S.R., Pelling, R.M., Hurley, K.C.
        1984, Solar hard X-ray microflares,
        {\it Astrophys. J.}, {\bf 283}, 421-425.
\ref Lin, R.P., Feffer,P.T., Schwartz,R.A. 2001,
        Solar Hard X-Ray Bursts and Electron Acceleration Down to 8 keV,
        {\it Astrophys. J.} {\bf 557}, L125-L128. 
\ref Lu, E.T., Hamilton, R.J., McTiernan, J.M., Bromund, K.R. 1993,
        Solar flares and avalanches in driven dissipative systems,
        {\it Astrophys. J.} {\bf 412}, 841-852. 
\ref Perez Enriquez, R. and Miroshnichenko, L.I. 1999,
        Frequency distributions of solar gamma ray events related
        and not related with SPEs 1989 -- 1995,
        {\it Solar Phys.}, {\bf 188}, 169-185. 
\ref Schwartz, R.A., Dennis, B.R., Fishman, G.J., Meegan, C.A., Wilson, R.B.,
        Paciesas, W.S. 1992,
        BATSE flare observations in solar cycle 22,
        in {\sl The Compton Observatory Science Workshop}, 
	C.R.Shrader, N.Gehrels, and B.R.Dennis (eds.), 
	NASA CP 3137 (NASA: Washington DC), p.457.
\ref Su, Y., Gan, W.Q., Li, Y.P. 2006,
        A statistical study of RHESSI flares,
        {\it Solar Phys.}, {\bf 238}, 61-72. 
\ref Tranquille, C., Hurley, K., Hudson, H. S. 2009,
        The Ulysses Catalog of Solar Hard X-Ray Flares,
        {\it Solar Phys.}, {\bf 258}, 141-166. 

\end{article}
\end{document}